\documentstyle[12pt,psfig,draft,lscape,array]{nature-ejb}

\def\simlt{\mathrel{\hbox{\rlap{\hbox{\lower4pt\hbox{$\sim$}}}\hbox{$<$}}}}
\def\simgt{\mathrel{\hbox{\rlap{\hbox{\lower4pt\hbox{$\sim$}}}\hbox{$>$}}}}

\def\ale{\mathrel{\hbox{\rlap{\hbox{\lower4pt\hbox{$\sim$}}}\hbox{$<$}}}}
\def\age{\mathrel{\hbox{\rlap{\hbox{\lower4pt\hbox{$\sim$}}}\hbox{$>$}}}}

\def\spose#1{\hbox to 0pt{#1\hss}}
\newcommand\lsim{\mathrel{\spose{\lower 3pt\hbox{$\mathchar"218$}}
     \raise 2.0pt\hbox{$\mathchar"13C$}}}
\newcommand\gsim{\mathrel{\spose{\lower 3pt\hbox{$\mathchar"218$}}
     \raise 2.0pt\hbox{$\mathchar"13E$}}}

\begin{document}
\noindent{ }\\
\noindent{\Large \bf Discovery and identification of the very high redshift afterglow of GRB~050904}\\

\noindent{\large J.~B.~Haislip$^1$, M.~C.~Nysewander$^1$, D.~E.~Reichart$^1$, A.~Levan$^2$, N.~Tanvir$^2$, S.~B.~Cenko$^3$, D.~B.~Fox$^4$, P.~A.~Price$^5$, A.~J.~Castro-Tirado$^6$, J.~Gorosabel$^6$, C.~R.~Evans$^1$, E.~Figueredo$^7$, C.~MacLeod$^1$, J.~Kirschbrown$^1$, M.~Jelinek$^6$, S.~Guziy$^6$, A.~de~Ugarte~Postigo$^6$, E.~S.~Cypriano$^7$, A.~LaCluyze$^1$, J.~Graham$^8$, R.~Priddey$^2$, R.~Chapman$^2$, J.~Rhoads$^9$, A.~S.~Fruchter$^9$, D.~Q.~Lamb$^{10}$, C.~Kouveliotou$^{11}$, R.~A.~M.~J.~Wijers$^{12}$, B.~P.~Schmidt$^{13}$, A.~M.~Soderberg$^{3}$, S.~R.~Kulkarni$^3$, F.~A.~Harrison$^{14}$, D.~S.~Moon$^3$, A.~Gal-Yam$^3$, M.~M.~Kasliwal$^{3}$, R.~Hudec$^{15}$, S.~Vitek$^{16}$, P.~Kubanek$^{17}$, J.~A.~Crain$^1$, A.~C.~Foster$^1$, M.~B.~Bayliss$^{1,10}$, J.~C.~Clemens$^1$, J.~W.~Bartelme$^1$, R.~Canterna$^{18}$, D.~H.~Hartmann$^{19}$, A.~A.~Henden$^{20}$, S.~Klose$^{21}$, H.-S.~Park$^{22}$, G.~G.~Williams$^{23}$, E.~Rol$^{24}$, P.~O'Brien$^{9}$, D.~Bersier$^{9}$, F.~Prada$^{6}$, S.~Pizarro$^7$, D.~Maturana$^7$, P.~Ugarte$^7$, A.~Alvarez$^7$,  A.~J.~M.~Fernandez$^6$, M.~J.~Jarvis$^{25}$, M.~Moles$^6$, E.~Alfaro$^6$, N.~D.~Kumar$^{1}$, C.~Mack$^{1}$, N.~Gehrels$^{26}$, S.~Barthelmy$^{26}$, D.~N.~Burrows$^4$}

\noindent{\small $^1$\,Department of Physics and Astronomy, University of North Carolina at Chapel Hill, Campus}

\vspace{-0.25 cm}
\noindent{\small Box 3255, Chapel Hill, NC 27599, USA}

\vspace{-0.25 cm}
\noindent{\small $^2$\,Centre for Astrophysics Research, University of Hertfordshire, College Lane, Hatfield, AL10}

\vspace{-0.25 cm}
\noindent{\small   9AB, United Kingdom}

\vspace{-0.25 cm}
\noindent{\small $^3$\,Department of Astronomy, California Institute of Technology, Pasadena, CA 91125, USA}

\vspace{-0.25 cm}
\noindent{\small $^4$\,Department of Astronomy and Astrophysics, 525 Davey Laboratory, Pennsylvania State }

\vspace{-0.25 cm}
\noindent{\small   University, University Park, PA 16802, USA}

\vspace{-0.25 cm}
\noindent{\small $^5$\,Institute for Astronomy, University of Hawaii, 2680 Woodlawn Drive, Honolulu, HI 96822,}

\vspace{-0.25 cm}
\noindent{\small USA}

\vspace{-0.25 cm}
\noindent{\small $^6$\,Instituto de Astrofisica de Andalucia, PO Box 3.004, 18.080 Granada, Spain}

\vspace{-0.25 cm}
\noindent{\small $^7$\,Southern Observatory for Astrophysical Research, Casilla 603, La Serena, Chile}

\vspace{-0.25 cm}
\noindent{\small $^8$\,Department of Astronomy, 601 Campbell Hall, University of California, Berkeley, CA 94720,}

\vspace{-0.25 cm}
\noindent{\small USA}

\vspace{-0.25 cm}
\noindent{\small $^9$\,Space Telescope Science Institute, 3700 San Martin Drive, Baltimore, MD 21218, USA}

\vspace{-0.25 cm} 
\noindent{\small $^{10}$\,Department of Astronomy and Astrophysics, University of Chicago, Chicago, IL 60637, USA}

\vspace{-0.25 cm}
\noindent{\small $^{11}$\,NASA Marshall Space Flight Center, National Space Science Technology Center, 320}

\vspace{-0.25 cm}
\noindent{\small   Sparkman Drive, Huntsville, AL 35805, USA}

\vspace{-0.25 cm}
\noindent{\small $^{12}$\,Astronomical Institute ``Anton Pannekoek'', University of Amsterdam and Center for High-}

\vspace{-0.25 cm}
\noindent{\small Energy Astrophysics, Kruislaan 403, 1098 SJ Amsterdam, Netherlands}

\vspace{-0.25 cm}
\noindent{\small $^{13}$\,Mount Stromlo and Siding Spring Observatories, Private Bag, Weston Creek P.O., Canberra}

\vspace{-0.25 cm}
\noindent{\small   ACT 2611, Australia}

\vspace{-0.25 cm}
\noindent{\small $^{14}$\,Space Radiation Laboratory, California Institute of Technology, MC 220-47, Pasadena, CA}

\vspace{-0.25 cm}
\noindent{\small   91125, USA}

\vspace{-0.25 cm}
\noindent{\small $^{15}$\,Astronomical Institute, Academy of Sciences of the Czech Republic, 25165 Ondrejov, Czech}

\vspace{-0.25 cm}  
\noindent{\small   Republic}

\vspace{-0.25 cm}
\noindent{\small $^{16}$\,Faculty of Electrotechnics, Czech Technical University,  121 35 Praha, Czech Republic}

\vspace{-0.25 cm}
\noindent{\small $^{17}$\,Integral Science Data Center, Chemin d'Ecogia 16, CH-1290 Versoix, Switzerland}

\vspace{-0.25 cm}
\noindent{\small $^{18}$\,Department of Physics and Astronomy, University of Wyoming, P.O. Box 3905, Laramie, WY}

\vspace{-0.25 cm}  
\noindent{\small   82072, USA}

\vspace{-0.25 cm}
\noindent{\small $^{19}$\,Clemson University, Department of Physics and Astronomy, Clemson, SC 29634, USA}

\vspace{-0.25 cm}
\noindent{\small $^{20}$\,American Association of Variable Star Observers, 25 Birch Street, Cambridge, MA 02138}

\vspace{-0.25 cm}
\noindent{\small USA}

\vspace{-0.25 cm}
\noindent{\small $^{21}$\,Thueringer Landessternwarte Tautenburg, Sternwarte 5, D-07778 Tautenburg, Germany}

\vspace{-0.25 cm}
\noindent{\small $^{22}$\,Lawrence Livermore National Laboratory, 7000 East Avenue, Livermore, CA 94550, USA}

\vspace{-0.25 cm}
\noindent{\small $^{23}$\,Multiple Mirror Telescope Observatory, University of Arizona, Tucson, AZ 85721, USA}

\vspace{-0.25 cm}
\noindent{\small $^{24}$\,Department of Physics and Astronomy, University of Leicester, Leicester LE1 7RH, United}

\vspace{-0.25 cm}
\noindent{\small Kingdom}

\vspace{-0.25 cm}
\noindent{\small $^{25}$\,Astrophysics, Department of Physics, University of Oxford, Keble Road, Oxford OX1 3RH,}

\vspace{-0.25 cm}
\noindent{\small   United Kingdom}

\vspace{-0.25 cm}
\noindent{\small $^{26}$\,NASA/Goddard Space Flight Center, Greenbelt, MD 20771, USA}\\

\date{\today}{} \headertitle{Discovery of very high redshift GRB}
\mainauthor{Haislip et al.}

\hrule 
\medskip
{\noindent \bf In 2000, Lamb and Reichart\cite{lr00} predicted that
$\gamma$-ray bursts (GRBs) and their afterglows occur in sufficient numbers and at sufficient brightnesses at very high
redshifts ($z > 5$) to eventually replace quasars as the preferred probe of element formation and
reionization in the early universe and to be used to characterize the star-formation history of the early universe, 
perhaps back to when the first stars formed (see also Refs.\,\cite{cl00}$^,$\cite{bl00}).  Here we report the
discovery of the afterglow of GRB~050904 and the identification of GRB~050904 as the first very high redshift GRB.\cite{hrc+05a}$^-$\cite{hnr+05}  We measure its
redshift to be $6.39^{+0.11}_{-0.12}$, which is consistent with the reported spectroscopic redshift ($6.29 \pm 0.01$)\cite{kyk+05}.  Furthermore, just redward of Ly$\alpha$ the flux is suppressed by a factor of three on the first night, but returns to expected levels by the fourth night.  We propose that this is due to absorption by
molecular hydrogen that was excited to rovibrational states by the GRB's
prompt emission, but was then overtaken by the jet.  Now that very high redshift GRBs have been shown to
exist, and at least in this case the afterglow was very bright, observing programs that are designed to capitalize on this science 
will likely drive a new era of study of 
the early universe, using GRBs as probes.
}
\medskip
\hrule
\bigskip
\bigskip
At 01:51:44 UT on September 4, 2005, Swift's Burst Alert Telescope (BAT)
detected GRB~050904 and 81 seconds later a 4$^\prime$-radius localization was
distributed to observers on the ground.  Swift's X-Ray Telescope (XRT)
automatically slewed to the BAT localization and 76 minutes after the
burst a 6$^{\prime\prime}$-radius XRT localization was distributed.\cite{cab+05}  We began remote
observations with the 4.1-m Southern Observatory for Astrophysical Research
(SOAR) telescope atop Cerro Pachon in Chile beginning 3.0 hours after the
burst.  Using the Ohio State InfraRed Imager/Spectrometer (OSIRIS) in
imaging mode, we discovered a relatively bright (J $\approx 17.4$ mag) and fading
near-infrared (NIR) source within the XRT localization (Fig.~1).\cite{hrc+05a}

Simultaneously, we also observed the XRT localization at visible
wavelengths with three telescopes:  one of the six 0.41-m Panchromatic
Robotic Optical Monitoring and Polarimetry Telescopes (PROMPT) that we are
building atop Cerro Tololo, which is only 10 km away from Cerro Pachon; the
60-inch telescope at Palomar Observatory in California; and the 3.5-m
telescope at Calar Alto Observatory in Spain.  None of these telescopes
detected the afterglow to relatively deep limiting magnitudes\cite{hrc+05a}$^,$\cite{fc05} (Fig.~1), nor did the 0.30-m Burst Observer and Optical Transient Exploring System (BOOTES) 1B telescope in El Arenosillo, Spain, which
began imaging the field only 2.1 minutes after the burst\cite{jcu+05}.  This
implied that the GRB either occurred at a very high redshift or that it was very heavily extinguished by dust.\cite{hrc+05a}

We then ruled out the extinction hypothesis by obtaining longer-wavelength
NIR observations with SOAR that same night:  In the NIR, the spectral index of the afterglow is 
$-1.25^{+0.15}_{-0.14}$; however, the spectral
index between NIR and visible wavelengths is steeper than $-5.9$ (Fig.~2).  This is too sharp of a transition to be explained by dust extinction.\cite{r01} 
Consequently, we identified GRB~050904 as a very high redshift GRB and
constrained its redshift to be $6 \lsim z \lsim 8$.\cite{hrc+05b}$^,$\cite{r05}  We then obtained
shorter-wavelength NIR detections with SOAR the following night, from which
we narrowed this range to its lower end:  $z \approx 6$.\cite{hnr+05}  Our
photometric redshift was later confirmed by two other groups: a photometric
redshift that was obtained with one of the 8.2-m Very Large Telescopes\cite{agd+05} and a spectroscopic redshift of $6.29 \pm 0.01$ that was obtained with
the 8.2-m Subaru telescope\cite{kyk+05}.

Our global monitoring campaign spanned four nights and also included the 8.1-m
Gemini South (Fig.~1), 3.8-m UKIRT, and 3.0-m IRTF telescopes (Tab.~1).  Between $\approx$3 hours and $\approx$0.5 days after the burst, the fading of the afterglow appears to be well described by a power law of index $-1.36^{+0.07}_{-0.06}$\,\cite{hrc+05b}$^,$\cite{r05} However, after this time the fading appears to have slowed to a temporal index of $-0.82^{+0.21}_{-0.08}$.\cite{hnr+05}$^,$\cite{dac+05a}$^,$\cite{dac+05b}  A single power law description is ruled out at the 3.7$\sigma$ credible level.  One possible explanation is that our initial SOAR observations caught the tail end of a reverse shock that had been stretched out in time by a factor of 7.29 due to cosmological time dilation.  Another possibility is that we are undersampling a light curve that is undergoing temporal variations, such as in the afterglows of GRBs 021004 and 030329.

Using all of our photometry except for a Z-band (0.84 -- 0.93 $\mu$m)
measurement from the first night (see below), we refine our earlier measurement of the redshift, again by assuming negligible emission blueward of Ly$\alpha$ (Fig.~2):  We measure $z = 6.39^{+0.11}_{-0.12}$, which is consistent with the reported spectroscopic redshift. 
For $H_0 = 71$ km s$^{-1}$ Mpc$^{-1}$, $\Omega_m=0.27$,
and $\Omega_\Lambda = 0.73$\cite{svp+03}, this corresponds to about 900
million years after the Big Bang, when the universe was about 6\% of its current age.  The next most distant GRB that has been identified occurred at $z = 4.50$\cite{ahp+00}, which was
about 500 million years later when the universe was about 10\% of its
current age.

The Z-band measurement, which was obtained with UKIRT 11 hours after the burst, is a 
factor of three below the fitted model (Fig.~3).  We have carefully modeled the spectral responses of the filter and of the detector and have checked the magnitudes of other sources in the field against similar z$^\prime$-band measurements that we obtained with Gemini South 3.2 days after the burst (Tab.~1).  This appears to be real and corresponds to a factor-of-three suppression of the flux between source-frame 1270 \AA\, and Ly$\alpha$ $\approx$0.5 days after the burst.  However, there was negligible suppression of the flux redward of source-frame 1600 \AA, which is again too
sharp of a transition to be explained by
dust extinction.\cite{r01}  Furthermore, by $\approx$3 days after the burst this signature had disappeared completely.  We propose that this is due to absorption by
molecular hydrogen that was excited to rovibrational states by the GRB's
prompt emission\cite{d00}, but then overtaken by the jet.  This implies a column density of molecular hydrogen along the line of sight of $>$$10^{18} - 10^{19}$ cm$^{-2}$\,\cite{d00} and a density of $\sim$$200$ cm$^{-3}$ (Fig.~3), which is consistent with a molecular cloud environment.\cite{rp02}  

One of the most exciting aspects of this discovery is the brightness of
the afterglow:  Extrapolating back to a few minutes after the burst, the
afterglow must have been exceptionally bright redward of Ly$\alpha$ for
the robotic 0.25-m TAROT telescope to detect it in unfiltered visible-light
observations.\cite{kba05}  Extrapolating our J-band light curve back to these times
yields J $\sim$ 11 -- 12 mag.  This suggests that by pairing visible-light
robotic telescopes with NIR robotic telescopes, and these with larger
telescopes that are capable of quick-response NIR spectroscopy, all
preferably at the same site, at least some very high redshift afterglows
will be discovered, identified, and their NIR spectrum taken while still
sufficiently bright to serve as an effective probe of the conditions of
the early universe.\cite{rnm+05}



\section*{Acknowledgments} 
D.E.R very gratefully acknowledges support from NSF's MRI, CAREER, PREST, and REU
programs, NASA's APRA, Swift GI and IDEAS programs, and especially Leonard Goodman and Henry Cox.  DER also very gratefully acknowledges Wayne Christiansen, Bruce Carney, and everyone who has worked to make SOAR a reality over the past 19 years.  A.L. and N.T. thank Brad Cavanagh and Andy Adamson of the JAC
for their speedy assistance in acquiring and reducing the
UKIRT WFCAM data.

\section*{Author Information}
Correspondence and requests for materials should be addressed to D.E.R (reichart@physics.unc.edu).


\clearpage

\begin{table}
\begin{center}
\begin{tabular}{>{\small}l >{\small}c >{\small}c >{\small}c >{\small}c}
\hline
\hline
\normalsize Date (UT) & \normalsize Mean $\Delta$t & \normalsize Filter & \normalsize Magnitude$^a$ & \normalsize Telescope \\
\hline


Sep 4.0795 & 2.80 min & R & $>$18.2 & 0.30-m BOOTES-1B \\
Sep 4.0821 & 6.46 min & R & $>$18.3 & 0.30-m BOOTES-1B \\
Sep 4.0868 & 13.22 min & R & $>$19.2 & 0.30-m BOOTES-1B \\
Sep 4.0956 & 25.95 min & R & $>$19.5 & 0.30-m BOOTES-1B \\
Sep 4.1151 & 53.96 min & R & $>$19.9 & 0.30-m BOOTES-1B \\
Sep 4.1535 & 109.30 min & R & $>$21.0 & 3.5-m Calar Alto \\
Sep 4.206 & 3.07 hr & J & 17.36 $\pm$ 0.04 & 4.1-m SOAR \\
Sep 4.213 & 3.25 hr & J & 17.35 $\pm$ 0.04 & 4.1-m SOAR \\
Sep 4.220 & 3.42 hr & J & 17.61 $\pm$ 0.04 & 4.1-m SOAR \\
Sep 4.248 & 4.08 hr & z & $>$18.8 & 60-inch Palomar \\
Sep 4.355 & 6.66 hr & R & $>$22.3 & 60-inch Palomar \\
Sep 4.366 & 6.91 hr & $^b$ & $>$20.1 & 0.41-m PROMPT-5 \\
Sep 4.390 & 7.49 hr & J & 18.66 $\pm$ 0.15 & 4.1-m SOAR \\
Sep 4.402 & 7.78 hr & K$_{\rm s}$ & 16.77 $\pm$ 0.07 & 4.1-m SOAR \\
Sep 4.416 & 8.12 hr & i & $>$21.1 & 60-inch Palomar \\
Sep 4.486 & 9.79 hr & H & 18.17 $\pm$ 0.06 & 3.8-m UKIRT \\
Sep 4.488 & 9.86 hr & J & 19.02 $\pm$ 0.06 & 3.8-m UKIRT \\
Sep 4.502 & 10.18 hr & K & 17.38 $\pm$ 0.06 & 3.8-m UKIRT \\
Sep 4.518 & 10.57 hr & K$^\prime$ & 17.55 $\pm$ 0.03 & 3.0-m IRTF \\
Sep 4.551 & 11.35 hr & Z & 22.08 $\pm$ 0.16 & 3.8-m UKIRT \\
Sep 4.565 & 11.69 hr & J & 19.25 $\pm$ 0.07 & 3.8-m UKIRT \\
Sep 5.198 & 26.90 hr & Y & 20.42 $\pm$ 0.26 & 4.1-m SOAR \\
Sep 5.246 & 28.03 hr & J & 20.16 $\pm$ 0.17 & 4.1-m SOAR \\
Sep 5.322 & 29.87 hr & I$_{\rm c}$ & $>$20.2 & 0.41-m PROMPT-3 \\
&&&& $+$ 0.41-m PROMPT-5\\
Sep 6.30 & 2.22 day & J & 20.60 $\pm$ 0.23 & 4.1-m SOAR \\
Sep 6.35 & 2.27 day & Y & 20.98 $\pm$ 0.34 & 4.1-m SOAR \\
Sep 7.21 & 3.13 day & i$^\prime$ & $>$25.4 & 8.1-m Gemini South \\
Sep 7.23 & 3.15 day & r$^\prime$ & $>$26.5 & 8.1-m Gemini South \\
Sep 7.24 & 3.16 day & z$^\prime$ & 23.36 $\pm$ 0.14 & 8.1-m Gemini South \\
\hline
$^a$ Upper limits are 3$\sigma$.\\
$^b$ Unfiltered, calibrated to R$_{\rm c}$.


\end{tabular}
\end{center}

\caption[]{Observations of the afterglow of GRB 050904.  We calibrated the i$^\prime$r$^\prime$z$^\prime$ measurements using stellar SDSS sources and derived R$_{\rm c}$I$_{\rm c}$ field calibrations from the SDSS field calibrations.  We obtained YJHK$_{\rm s}$K field calibrations using SOAR and ZJHK field calibrations using UKIRT.  The JHK field calibrations are in agreement with each other and with 2MASS.  The UKIRT WFCAM Z bandpass was designed to match the effective wavelength of the SDSS z$^\prime$ bandpass (0.876 vs.~0.887 $\mu$m), but with a rectangular profile.  The standard deviation of the magnitude differences for all stellar SDSS sources in the UKIRT Z and Gemini South z$^\prime$ fields is only 0.064 mag.  When converting from magnitudes to spectral fluxes, we used the correct zero points for Z and z$^\prime$, respectively.  When fitting to these spectral fluxes, we used the actual UKIRT WFCAM Z and Gemini South GMOS-S z$^\prime$ bandpasses.  Consequently, we conclude that the factor of three deficit that we measure in the Z band relative to the fitted model 11 hours after the burst is real.
}
\end{table}

\clearpage

\begin{figure}
\centerline{\psfig{file=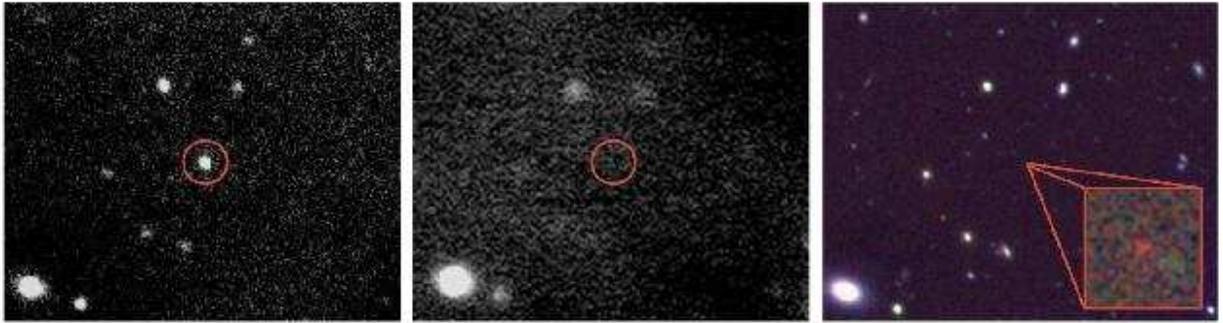,width=6.5in}}
\caption[]{{\it Left panel:}  NIR discovery image of the bright (J $= 17.36 \pm 0.04$ mag)
afterglow of GRB~050904 from 4.1-m SOAR atop Cerro Pachon, Chile.  {\it Middle
panel:}  Near-simultaneous non-detection of the afterglow at visible
wavelengths (unfiltered, calibrated to R$_{\rm c} > 20.1$ mag) from one of the six
0.41-m PROMPT telescopes that we are building atop Cerro Tololo, which is only 10 km away from Cerro Pachon.  {\it Right panel:}  Color composite (riz) image of the afterglow 3.2 days after the burst from 8.1-m Gemini South, which is also atop Cerro Pachon.}
\end{figure}

\clearpage

\begin{figure}
\centerline{\psfig{file=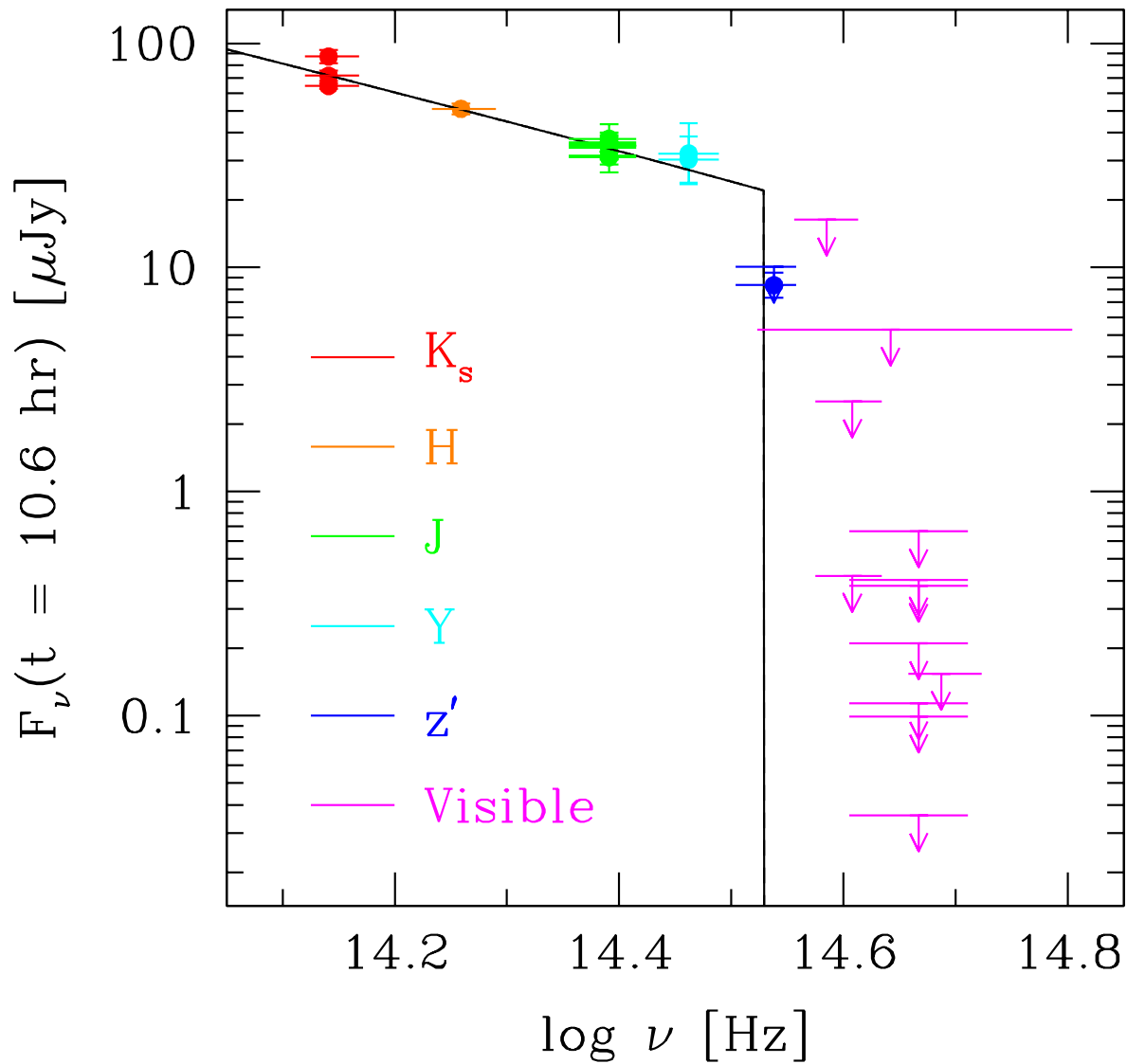,width=6.5in}}
\caption[]{Spectral flux distribution of the afterglow of GRB~050904
scaled to 10.6 hours after the burst and our best-fit model:  a power-law spectrum with negligible emission blueward of Ly$\alpha$.  Shallower power-law spectra can be obtained with the addition of source-frame dust, but this alone cannot explain the sharp drop in and blueward of the z$^\prime$ band.\cite{r01}  We 
take Galactic $E(B-V) = 0.060$ mag\cite{sfd98}.  
}
\end{figure}

\clearpage

\begin{figure}
\centerline{\psfig{file=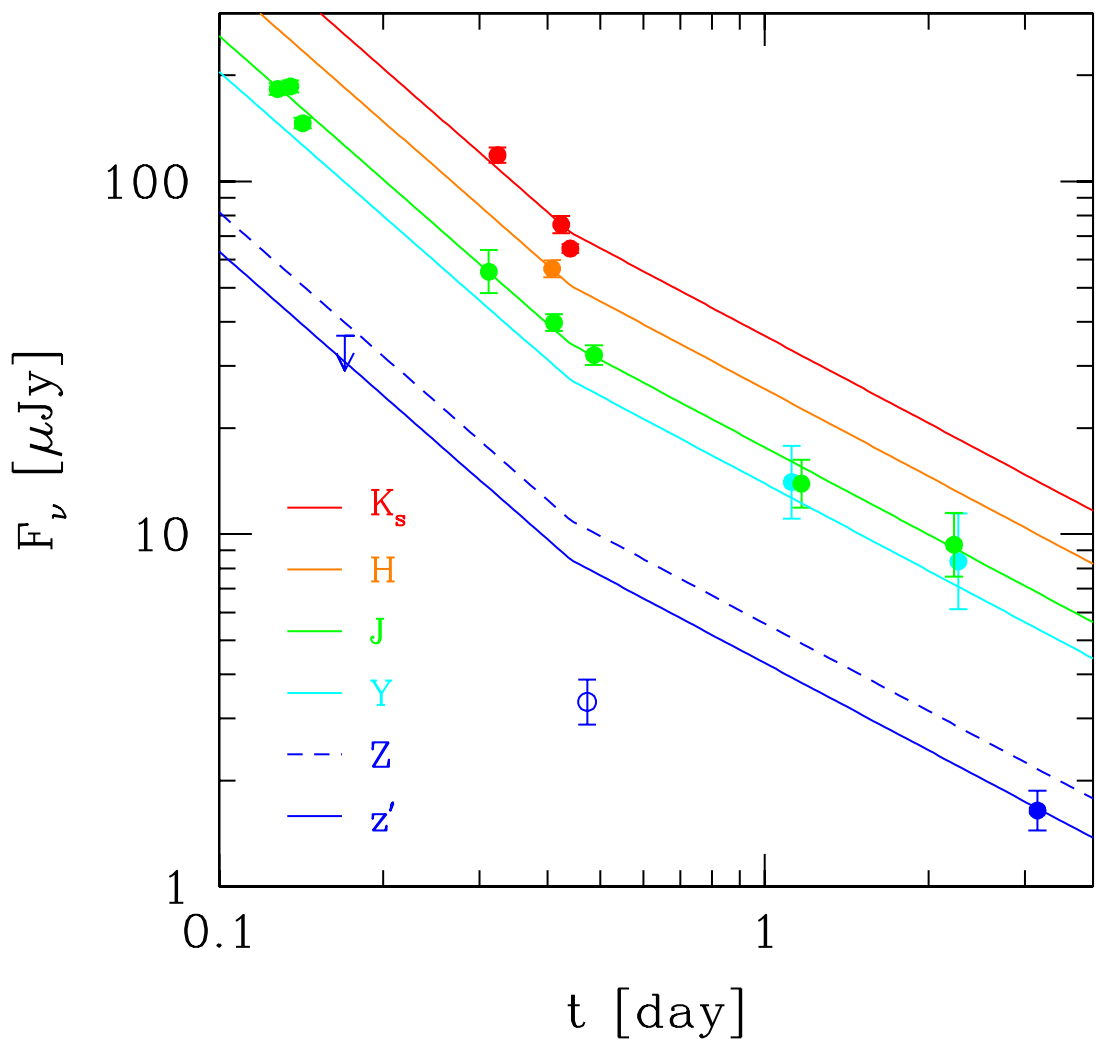,width=6in}}
\caption[]{NIR and z$^\prime$-band light curves of the afterglow of GRB~050904 and
our best-fit model.  A single power law description is ruled out at the 3.7$\sigma$ credible level.  
Following the formalism of Frail et al.\cite{fks+01}, given GRB~050904's redshift and fluence\cite{sbb+05} the non-detection of a jet break in the light curve prior to 3.2 days after
the burst implies that the opening/viewing angle of the jet is
$\gsim$2.6$^\circ$ and that the total energy that was released in $\gamma$ rays is $\gsim$$4.0\times10^{50}$ erg.
Finally, the Z-band measurement (unfilled circle) is a factor of three
below the fitted model, but this appears to be real.  
If due to absorption by excited molecular hydrogen as we propose, in this wavelength range a factor-of-three supression of the flux implies that the column density of excited molecular hydrogen in front of the jet at $\approx$11 hours after the burst in the observer frame is $N(H_2) \sim 3 \times 10^{18}$ cm$^{-2}$\cite{d00}.  Consequently, this absorption signature should disappear by the time that the jet travels an additional distance $N(H_2) / n(H_2)$.  The distance that the jet travels as a function of observer-frame time $t_{obs}$ is given by $2\gamma^2(t_{obs})ct_{obs}/(1+z)$, where for GRB~050904 $\gamma(t_{obs}) \approx 30n^{-1/8}(t_{obs}/11\,{\rm hr})^{-3/8}$\,\cite{spn98}$^,$\cite{sbb+05} and $n$ is the overall density of the medium.  Consequently, the jet should overtake the region of excited molecular hydrogen at $t_{obs} \sim 11[1+10n^{1/4}/n(H_2)]^4$ hours.  For $n^{4/3}(H_2)/n^{1/3} \sim 200$, $t_{obs} \sim 1$ day after the burst.  Absorption by excited molecular hydrogen might also have been seen in the afterglow of GRB~980329.\cite{rlm+99}$^,$\cite{f99}$^,$\cite{d00}

}
\end{figure}


\end{document}